\documentclass{MYws-ijmpa}
\usepackage{overcite}

\def\Journal#1#2#3#4{{#1} {\bf #2}, #3 (#4)}
\def\refjl#1#2#3#4#5#6{\bibitem{#1} #2, \Journal{#3}{#4}{#6}{#5}.}

\def\PRL{\em Phys. Rev. Lett.}

\def\NP{\em Nucl. Phys.}

\def\PL{\em Phys. Lett.}
\def\PR{\em Phys. Rev.}

\def\RPP{\em Rep. Prog. Phys.}

\def\PPNP{\em Prog. Part. Nucl. Phys.}

\newcommand{\eqn}[1]{(\ref{#1})}
\newcommand{\be}{\begin{equation}}
\newcommand{\ee}{\end{equation}}
\newcommand{\no}{\nonumber}
\newcommand{\bel}[1]{\be\label{#1}}
\newcommand{\ba}{\begin{array}{c}}
\newcommand{\bat}{\begin{array}{cc}}
\newcommand{\ea}{\end{array}}
\newcommand{\beqn}{\begin{eqnarray}}
\newcommand{\eeqn}{\end{eqnarray}}

\def\gap{\;\lower3pt\hbox{$\buildrel > \over \sim$}\;}
\def\lap{\;\lower3pt\hbox{$\buildrel < \over \sim$}\;}
\newcommand{\cL}{{\cal L}}

\newcommand{\cM}{{\cal M}}

\begin{document}

\markboth{A. Pich}{Chiral Perturbation Theory}

\catchline{}{}{}{}{}

\title{PRESENT STATUS OF CHIRAL PERTURBATION THEORY\footnote{
Invited talk at the $10^{\mathrm{th}}$ International Symposium on Meson--Nucleon
Physics and the Structure of the Nucleon
(MENU 2004), Beijing, China, August 29 -- September 4, 2004}}

\author{A. PICH}
\address{Departament de F\'{\i}sica Te\`orica, IFIC,
Universitat de Val\`encia--CSIC,\\
Apt. Correus 22085,
E-46071 Val\`encia, Spain. \ E-mail: Antonio.Pich@ific.uv.es}

\maketitle


\begin{abstract}
The basic ideas and methods of chiral perturbation theory are
briefly reviewed. I discuss the recent attempts to build an
effective Lagrangian in the resonance region and summarize
the known large--$N_C$ constraints on the low-energy chiral couplings.
\end{abstract}



\section{Chiral Symmetry}
\label{sec:chpt}

With $n_f$ massless quark flavours, the QCD Lagrangian
is invariant under global \ $SU(n_f)_L\otimes SU(n_f)_R$
transformations of the left- and right-handed quarks in flavour space.
The symmetry group spontaneously breaks down to the diagonal
subgroup $SU(n_f)_{L+R}$ \ and \
$n_f^2-1$ pseudoscalar massless Goldstone bosons appear in the theory,
which for $n_f=3$ can be identified with the
eight lightest hadronic states $\phi^a =\{\pi$, $K$, $\eta\}$.
These pseudoscalar fields are usually parameterized through the
$3\times 3$ unitary matrix \
$U(\phi) =  u(\phi)^2 =\exp{\left\{i\lambda^a\phi^a/f\right\}}$.

The Goldstone nature of the pseudoscalar mesons implies strong
constraints on their interactions, which can be most easily analyzed
on the basis of an effective Lagrangian.\cite{WE:79}
Since there is a mass gap separating the pseudoscalar octet from the
rest of the hadronic spectrum, we can build an effective field
theory containing only the Goldstone modes.\cite{EC:95,PI:95}
The low-energy effective Lagrangian can be organized in
terms of increasing powers of momenta (derivatives): \
$\cL = ?\sum_n \cL_{2n}$.

It is convenient to consider an extended QCD Lagrangian, with quark
currents coupled to external Hermitian matrix-valued sources
$l_\mu$, $r_\mu$, $s$, $p$. In addition to generate the QCD
Green functions, the external fields can be used to incorporate the
electromagnetic and semileptonic weak interactions, and the
explicit breaking of chiral symmetry through the quark masses:
$s = \cM + \ldots\, ,\,\cM= \hbox{\rm diag}(m_u,m_d,m_s)$.

At lowest order in derivatives and quark masses,
the most general effective Lagrangian
consistent with chiral symmetry has the form:\cite{GL:85}
\bel{eq:lowestorder}
\cL_2 = {f^2\over 4}\,
\langle D_\mu U^\dagger D^\mu U \, + \, U^\dagger\chi  \,
+  \,\chi^\dagger U\rangle\, ,
\qquad\qquad
\chi \equiv 2 B_0 \, (s + i p) \, ,
\ee
where \
$D_\mu U = \partial_\mu U -i r_\mu U + i U\, l_\mu$ ,
$\langle A\rangle$ denotes the flavour trace of the matrix $A$
and $B_0$ is a constant, which, like $f$, is not fixed by
symmetry requirements alone.
One finds that $f$ equals the pion decay constant
(at lowest order) $f = f_\pi = 92.4$~MeV, while $B_0$ is
related to the quark condensate:
\bel{eq:B0}
B_0 = -{\langle\bar q q \rangle\over f^2} =
{M_\pi^2\over m_u + m_d} = {M_{K^0}^2\over m_s + m_d} =
{M_{K^\pm}^2\over m_s + m_u}\, .
\ee
With only two low-energy constants, the lowest-order chiral
Lagrangian $\cL_2$ encodes in a very compact way all the
Current Algebra results obtained in the sixties.

The symmetry constraints become less powerful at higher orders.
At $O(p^4)$ we need ten additional coupling constants $L_i$
to determine the low-energy behaviour of the Green functions:\cite{GL:85}
\bel{eq:l4}
\cL_4  = L_1 \,\langle D_\mu U^\dagger D^\mu U\rangle^2 \, + \,
L_2 \,\langle D_\mu U^\dagger D_\nu U\rangle\,
  \langle D^\mu U^\dagger D^\nu U\rangle\, + \,\ldots
\ee
One-loop graphs with the lowest-order Lagrangian $\cL_2$ contribute
also at $O(p^4)$.
Their divergent parts are renormalized by the $\cL_4$ couplings,
which introduces a renormalization scale dependence.
The chiral loops generate non-polynomial contributions,
with logarithms and threshold factors as required by unitarity,
which are completely determined as functions of $f$ and the Goldstone
masses.

\begin{table}[tb]
\tbl{Phenomenological values of the renormalized couplings
$L_i^r(M_\rho)$ in units of $10^{-3}$.
The large--$N_C$ predictions obtained within the single-resonance
approximation are given in the last column.\protect\cite{EC:95,PI:95}}{
\label{tab:Lcouplings}
\renewcommand{\arraystretch}{1.1}
\begin{tabular}{|c|c|c|c|c|}
\hline
$i$ & \raisebox{0pt}[10pt][5pt]{$L_i^r(M_\rho)$} & $O(N_C)$  & Source &
$L_{i}^{N_C\to\infty}$
\\
\hline
\raisebox{0pt}[10pt]{$2L_1-L_2$}
& $-0.6\pm0.6$ & $O(1)$ &
$K_{e4}$, $\pi\pi\to\pi\pi$ & 0
\\
$L_2$ & $\hphantom{-}1.4\pm0.3$ & $O(N_C)$ &
$K_{e4}$, $\pi\pi\to\pi\pi$ & $\phantom{-}1.8$
\\
$L_3$ & $-3.5\pm1.1$ & $O(N_C)$ & $K_{e4}$, $\pi\pi\to\pi\pi$ & $-4.3$
\\
$L_4$ & $-0.3\pm0.5$ & $O(1)$ & Zweig rule & 0
\\
$L_5$ & $\hphantom{-}1.4\pm0.5$ & $O(N_C)$ & $F_K : F_\pi$ & $\phantom{-}2.1$
\\
$L_6$ & $-0.2\pm0.3$ & $O(1)$ & Zweig rule & 0
\\
$L_7$ & $-0.4\pm0.2$ & $O(1)$ & GMO, $L_5$, $L_8$ & $-0.3$
\\
$L_8$ & $\hphantom{-}0.9\pm0.3$ & $O(N_C)$ & $M_\phi$, $L_5$ & $\phantom{-}0.8$
\\
$L_9$ & $\hphantom{-}6.9\pm0.7$ & $O(N_C)$ &
$\langle r^2\rangle^\pi_V$ & $\phantom{-}7.1$
\\
$L_{10}$ & $-5.5\pm0.7$ & $O(N_C)$ & $\pi\to e\nu\gamma$ & $-5.4$
\\[1pt]\hline
\end{tabular}}
\end{table}

The precision required in present phenomenological applications
makes necessary to include corrections of $O(p^6)$.
This involves contributions from $\cL_4$ at one-loop and
$\cL_2$ at two-loops, which can be fully predicted.\cite{bce00}
However, the $O(p^6)$ chiral Lagrangian $\cL_6$ contains
90 (23) independent local terms of even (odd)
intrinsic parity.\cite{bce00,BGP:01}
The huge number of unknown couplings limits
the achievable accuracy.
Clearly, further progress will depend on our ability
to estimate these chiral couplings, which encode the
underlying QCD dynamics.

Several two-loop calculations have been already
performed.\cite{colangelo,bijnens,BGS:94,KMS:96,PS:02,kambor}
Thus, the non-local $O(p^6)$ contributions (chiral logarithms)
to many observables are known and, in some cases,
the local ambiguities can be reduced to a few subtraction constants
using dispersion relation techniques.

\section{Resonance Chiral Theory}
\label{sec:RChT}

The limit of an infinite number of quark colours
is a very useful starting point to understand many
features of QCD.\cite{HO:74,WI:79}
Assuming confinement,
the strong dynamics at $N_C\to\infty$ is given
by tree diagrams with infinite sums of hadron exchanges,
which correspond to the tree approximation to some local
effective Lagrangian. Hadronic loops generate corrections
suppressed by factors of $1/N_C$.
At $N_C\to\infty$, QCD has a larger symmetry
$U(3)_L\otimes U(3)_R\to U(3)_{L+R}$, and
one needs to include in the matrix $U(\phi)$ a ninth
Goldstone boson field, the $\eta_1$.
Resonance chiral theory provides
the correct framework to incorporate the massive mesonic
states.\cite{PI:02}

Let us consider a chiral-invariant Lagrangian
$\cL_R$, describing the couplings of resonance nonet multiplets
$V_i^{\mu\nu}(1^{--})$, $A_i^{\mu\nu}(1^{++})$, $S_i(0^{++})$ and $P_i(0^{-+})$ to
the Goldstone bosons:\cite{EGPR:89}
\beqn\label{eq:L_R}
\cL_R &=& \sum_i\;\biggl\{ {F_{V_i}\over 2\sqrt{2}}\;
\langle V_i^{\mu\nu} f_{+ \, \mu\nu}\rangle\, +\,
{i\, G_{V_i}\over \sqrt{2}} \,\,\langle V_i^{\mu\nu} u_\mu u_\nu\rangle
\, +\,
{F_{A_i}\over 2\sqrt{2}} \;\langle A_i^{\mu\nu} f_{- \, \mu\nu} \rangle
\biggr.\no\\ &&\hskip .5cm\biggl.\mbox{} +\,
c_{d_i} \; \langle S_i\, u^\mu u_\mu\rangle
\, +\, c_{m_i} \; \langle S_i\, \chi_+ \rangle
\, +\, i\, d_{m_i}\;\langle P_i\, \chi_- \rangle
\biggr\}\, ,
\eeqn
where \
$u_\mu \equiv i\, u^\dagger D_\mu U u^\dagger$, \
$f^{\mu\nu}_\pm\equiv u F_L^{\mu\nu} u^\dagger\pm  u^\dagger F_R^{\mu\nu} u$
\ with $F^{\mu\nu}_{L,R}$ the field-strength tensors of the
$l^\mu$ and $r^\mu$ flavour fields
\ and \
$\chi_\pm\equiv u^\dagger\chi u^\dagger\pm u\chi^\dagger u$.
The resonance couplings
$F_{V_i}$, $G_{V_i}$, $F_{A_i}$, $c_{d_i}$, $c_{m_i}$ and $d_{m_i}$
are of \ $O\left(\sqrt{N_C}\,\right)$.

The lightest resonances have an important impact on the
low-energy dynamics of the pseudoscalar bosons.
Below the resonance mass scale, the singularity associated with the
pole of a resonance propagator is replaced by the corresponding
momentum expansion; therefore, the exchange of virtual resonances generates
derivative Goldstone couplings proportional to powers of $1/M_R^2$.
At lowest order in derivatives, this gives the large--$N_C$ predictions
for the $O(p^4)$ couplings of chiral perturbation theory:\cite{EGPR:89}
\beqn\label{eq:vmd_results}
2\, L_1 = L_2 = \sum_i\; {G_{V_i}^2\over 4\, M_{V_i}^2}\, , & \qquad &
L_3 = \sum_i\;\left\{ -{3\, G_{V_i}^2\over 4\, M_{V_i}^2} +
{c_{d_i}^2\over 2\, M_{S_i}^2}\right\} \, ,
\no\\
L_5 = \sum_i\; {c_{d_i}\, c_{m_i}\over M_{S_i}^2} \, ,\hskip .89cm &&
L_8 = \sum_i\;\left\{ {c_{m_i}^2\over 2\, M_{S_i}^2} -
{d_{m_i}^2\over 2\, M_{P_i}^2}\right\} \, ,
\\
L_9 = \sum_i\; {F_{V_i}\, G_{V_i}\over 2\, M_{V_i}^2}\, ,\hskip .77cm &&
L_{10} = \sum_i\;\left\{ {F_{A_i}^2\over 4\, M_{A_i}^2}
 - {F_{V_i}^2\over 4\, M_{V_i}^2}\right\}  \, .
\no\eeqn
All these couplings are of $O(N_C)$, in agreement with the
counting indicated in Table~\ref{tab:Lcouplings}, while for the
couplings of $O(1)$ we get
$2\, L_1-L_2 = L_4 = L_6 = L_7 = 0$.

Owing to the $U(1)_A$ anomaly, the $\eta_1$ field is massive and it is often
integrated out from the low-energy chiral theory. In that case,
the $SU(3)_L\otimes SU(3)_R$ chiral coupling $L_7$ gets a contribution
from $\eta_1$ exchange:\cite{GL:85,EGPR:89}
\bel{eq:L7}
L_7 = - {\tilde{d}_m^2\over 2\, M^2_{\eta_1}} \, ,
\qquad\qquad\qquad
\tilde{d}_m = -{f\over\sqrt{24}} \, .
\ee

\section{Short-Distance Constraints}

The short-distance properties of the underlying QCD dynamics
impose some constraints on the low-energy
parameters.\cite{PI:02,EGLPR:89}
At leading order in $1/N_C$, the two-Goldstone matrix element of the
vector current, is characterized by
\bel{eq:VFF}
F_V(t)\, =\, 1\, + \, \sum_i\,
{F_{V_i}\, G_{V_i}\over f^2}\; {t\over M_{V_i}^2-t} \, .
\ee
Since the vector form factor $F_V(t)$ should vanish at infinite momentum
transfer $t$, the resonance couplings should satisfy
\bel{eq:SD1}
\sum_i\, F_{V_i}\, G_{V_i}\, =\, f^2\, .
\ee
Similarly,
the matrix element of the axial current between one Goldstone and
one photon is parameterized by the so-called axial form factor $G_A(t)$,
which vanishes at $t\to\infty$ provided that
\bel{eq:SD2}
\sum_i\,
\left(2\, F_{V_i}\, G_{V_i}- F_{V_i}^2\right) / M_{V_i}^2
\, =\, 0\, .
\ee
Requiring the scalar form factor $F^S(t)$, which governs the
two-pseudoscalar matrix element of the scalar quark current,
to vanish at $t\to\infty$, one gets the constraints:\cite{JOP:02}
\bel{eq:SD4}
4\,\sum_i\,c_{d_i}\, c_{m_i} = f^2 \, ,
\qquad\qquad
\sum_i\,  c_{m_i}\,\left( c_{m_i}-c_{d_i}\right) / M_{S_i}^2 = 0 \, .
\ee

Since gluonic interactions preserve chirality,
the two-point function built from a left-handed and a right-handed
vector quark currents
$\Pi_{LR}(t)$ satisfies an unsubtracted dispersion relation.
In the chiral limit, it vanishes faster than $1/t^2$
when $t\to\infty$; this implies the well-known Weinberg conditions:\cite{WE:67}
\bel{eq:SD3}
\sum_i\,\left( F_{V_i}^2 - F_{A_i}^2\right) = f^2 \, ,
\qquad\qquad
\sum_i\,\left( M_{V_i}^2 F_{V_i}^2 - M_{A_i}^2 F_{A_i}^2\right)
= 0 \, .
\ee

The two-point correlators of two scalar or two pseudoscalar
currents would be equal if chirality was preserved.
For massless quarks, $\Pi_{SS-PP}(t)$ vanishes as $1/t^2$ when
$t\to\infty$, with a coefficient proportional to  
$\alpha_s\,\langle\bar q\Gamma q\,\bar q\Gamma q\rangle
\sim\alpha_s\,\langle\bar q q\rangle^2 \sim \alpha_s\, B_0^2$.
Imposing this behaviour, one gets:\cite{GP:00}
\bel{eq:SD5}
8\,\sum_i\left( c_{m_i}^2 - d_{m_i}^2\right) = f^2  ,
\qquad\qquad
\sum_i\left( c_{m_i}^2 M_{S_i}^2 - d_{m_i}^2 M_{P_i}^2\right) =
3\,\pi\alpha_s\, f^4 /4\, .
\ee

\section{Single-Resonance Approximation}

Let us approximate each infinite resonance sum with
the first meson nonet contribution.
This is meaningful at
low energies where the contributions from higher-mass states are
suppressed by their corresponding propagators.
The resulting short-distance constraints are
matching conditions between an effective theory
below the scale of the second resonance multiplets
and the underlying QCD dynamics.

With this approximation, Eqs.~\eqn{eq:SD1}, \eqn{eq:SD2} and \eqn{eq:SD3}
determine the vector and axial-vector couplings in terms of $M_V$
and $f$:\cite{EGLPR:89}
\bel{eq:VA_coup}
F_V = 2\, G_V = \sqrt{2}\, F_A = \sqrt{2}\, f \, ,
\qquad\qquad
M_A = \sqrt{2}\, M_V \, .
\ee
The scalar\cite{JOP:02}
and pseudoscalar parameters are obtained
from \eqn{eq:SD4} and \eqn{eq:SD5}:\cite{PI:02}
\bel{eq:SP_coup}
c_m = c_d = \sqrt{2}\, d_m = f/2 \, ,
\qquad\qquad
M_P = \sqrt{2}\, M_S \, \left(1 - \delta\right)^{1/2}\, .
\ee
The last relation involves a small correction \
$\delta \approx 3\,\pi\alpha_s f^2/M_S^2 \sim 0.08\,\alpha_s$,
which we can neglect together with the tiny effects from
light quark masses.

Inserting these predictions into Eqs.~\eqn{eq:vmd_results},
one finally gets all $O(N_C\, p^4)$ chiral perturbation theory
couplings, in terms of $M_V$, $M_S$ and $f$:
\bel{eq:Li_SRA_1}
2\, L_1 = L_2 = \frac{1}{4}\, L_9 = -\frac{1}{3}\, L_{10}
= {f^2\over 8\, M_V^2}\, ,
\ee
\bel{eq:Li_SRA_2}
L_3 = -{3\, f^2\over 8\, M_V^2} + {f^2\over 8\, M_S^2}\, ,
\qquad\quad
L_5 ={f^2\over 4\, M_S^2}\, ,
\qquad\quad
L_8 = {3\, f^2\over 32\, M_S^2}\, .
\ee
The last column in Table~\ref{tab:Lcouplings} shows the
results obtained with $M_V = 0.77$~GeV,
$M_S = 1.0$~GeV and $f=92$~MeV. Also shown is the $L_7$
prediction in \eqn{eq:L7}, taking
$M_{\eta_1} = 0.80$~GeV. The agreement with
the measured values is a clear
success of the large--$N_C$ approximation.
It demonstrates that the lightest resonance multiplets
give indeed the dominant effects at low energies.

The study of other Green functions provides further matching
conditions between the hadronic and fundamental QCD descriptions.
Clearly, it is not possible to satisfy all of them
within the single-resonance approximation. A useful generalization is the
so-called {\it Minimal Hadronic Ansatz}, which
keeps the minimum number of resonances compatible with all known
short-distance constraints for the problem at hand.\cite{KPdR}
Some $O(p^6)$ chiral couplings have been already
analyzed in this way, by studying an appropriate set of three-point
functions.\cite{MO:95,KN:01,RPP:03,BGLP:03}

\section{Subleading $1/N_C$ Corrections}

The large--$N_C$ limit provides a very successful description
of the low-energy dynamics.\cite{PI:02}
However, we are still lacking a systematic procedure to incorporate
next-to-leading contributions in the $1/N_C$ counting.
Up to now, the effort has concentrated in pinning down the most relevant
subleading effects, such as the resonance widths which regulate the corresponding poles
in the meson propagators,\cite{GP:97} or the role of final state interactions in
the physical amplitudes.\cite{JOP:02,GP:97,PaP:01,IAM}

Quantum loops including virtual resonance propagators constitute a
major technical challenge.\cite{CP:01,RSP:04}
Their ultraviolet divergences require higher dimensional
counterterms, which could generate a problematic behaviour at large momenta.
Thus, it is necessary to investigate
the short-distance QCD constraints at the next-to-leading order in $1/N_C$.
A first step in this direction is the recent one-loop calculation of the vector
form factor in the resonance chiral theory.\cite{RSP:04}
Further work towards a more formal renormalization procedure is in progress.


\section*{Acknowledgements}

This work has been supported 
by EU HPRN-CT2002-00311 (EURIDICE),  MCYT
(FPA-2001-3031) and Generalitat Valenciana
(GRUPOS03/013, GV04B-594).


\end{document}